\documentclass{article}

\usepackage{arxiv}

\usepackage[utf8]{inputenc} 
\usepackage[T1]{fontenc}    
\usepackage{hyperref}     
\usepackage{url}            
\usepackage{booktabs}       
\usepackage{amsfonts}      
\usepackage{nicefrac}       
\usepackage{microtype}   
\usepackage{lipsum}		
\usepackage{graphicx}
\usepackage{doi}
\usepackage{authblk}
\usepackage{mathtools}
\usepackage{amsmath}
\usepackage[thinc]{esdiff}
\usepackage{xcolor}
\usepackage{pdfpages}
\usepackage[export]{adjustbox}
\usepackage[
style=chem-acs
]{biblatex}
\title{Impact of surfactant and flow rate on the electrical properties of activated carbon black suspensions}

\date{}

\renewcommand{\headeright}{ }
\renewcommand{\undertitle}{ }
\renewcommand{\shorttitle}{ }

\author{KangJin Lee}
\author{Jesse S. Wainright}
\author{Christopher L. Wirth}

\affil{Department of Chemical and Biomolecular Engineering, Case Western Reserve University, \\ 10900 Euclid Ave, Cleveland, Ohio 44106, USA}

\begin{document}
\maketitle

\begin{abstract}
Carbon black slurries are a key component in redox flow batteries as the large surface area provided by the particles allows an increase in the battery capacity without facing limitations posed by many solid-state batteries such as safety hazard or cost. However, these conductive slurries often have complex mechanical and electrical responses because of the heterogeneous nature of the suspensions. Utilization of these slurries in a redox flow battery requires understanding of how additives impact the material responses and associated battery performance. This work focuses on the electrochemical performance of the slurry at flow rate conditions matching those of battery operation. In addition, the impact of a nonionic surfactant (Triton X-100) on the conductivity and capacitance of the slurry was measured. Experimental results show that the full capacitive contribution of the carbon black particles can only be measured at low flow rates and low scan rates, while the conductive contribution can be measured at all scan rates in flowing conditions. Upon the addition of surfactants, there is a gradual decrease in the electrochemical performance with increasing surfactant concentration until the surface of the carbon black particle is saturated. Once saturated, the conductive carbon particles no longer contribute to the electronic conductivity of the slurry. Results presented herein on the electrochemical response of the slurry to the addition of surfactant are in stark contrast to the mechanical response. While previous work has shown a smaller change in the response, followed by a step-change at a critical surfactant concentration, electrochemical data shows a gradual transition. Comparison of these behaviors suggests a difference in the mechanisms for how mechanical networks form in comparison to charge transfer networks for this particular slurry chemistry.
\end{abstract}


\section{Introduction}
 
Redox Flow Batteries (RFBs) are one of the alternative technologies to solid-state batteries that have the potential to be used for long-term energy storage at the grid level\cite{park2016material, dunn2011electrical, weber2011redox}. RFBs decouple and store electrolytes in external reservoirs, allowing easier scale-up without having to face fire hazards or storage space limitations commonly associated with solid-state batteries\cite{park2016material}. In addition, RFBs' flexibility in design, adaptability to new chemistry, and lower cost of operation and maintenance have shown that it can be a promising solution for long-term energy storage\cite{de2006redox}.\\

Currently, there are many different types of RFBs with varying chemistry such as all-vanadium, Fe/Cr, Zn/Br, and all-iron couple chemistry\cite{park2016material,de2006redox}. These chemistries are typically limited not only by their performance but also by the material cost and availability. Considering that the use of a single species is more advantageous since it can be readily dealt with and electrolytes can be mixed when being recycled, an all-iron flow battery becomes one of the highly favorable batteries as iron is less costly compared to vanadium or other commonly used metals\cite{hawthorne2014studies}. Even in all-iron flow batteries; however, a major challenge exists where the iron metal can plate onto the current collector, leading to poor performance in hybrid configuration\cite{tam2023electrochemical}. To overcome this challenge, many RFBs have started to use slurry electrodes as a solution by dispersing conductive particles in the electrolytes\cite{petek2015slurry}. With conductive particles in the electrolyte, the metal iron can plate onto the particles instead of the current collector, and the particles can be cycled out of the cell and throughout the system, leaving minimal negative impact and also easily allowing scalability. \\

Carbon black (CB) particles have shown their advantages in being a key component in formulating slurry electrodes as they are not only low in cost but have inherent chemical resistances and can form electrical percolation at relatively low loadings\cite{youssry2013non, richards2017clustering, chakrabarti2014application}. More specifically, activated CB particles have high surface area and porosity which are both advantageous properties for the battery\cite{boehm1994some}. The large surface area of the particles can provide more sites for reduced metals to deposit, increasing the energy storage capacity of the battery even at low particle loadings. The particular type of surface modification plays an important role in the behavior. For example, hydrophobic activated CB particles tend to form aggregates in aqueous electrolyte medium due to van der Waals forces, especially at lower pH levels where the battery operates\cite{parant2017flowing, hawthorne2014studies}. CB particle aggregation in these systems cause particles to form networks that span across the system, creating a viscoelastic gel with measurable yield stress\cite{das2024surfactant, helal2016simultaneous, zaccarelli2007colloidal}. When these networks form, they can negatively impact the slurry stability and cause phase separation through sedimentation\cite{lee2023surfactant} and significantly increase the slurry viscosity resulting in clogging of the flow battery channels\cite{presser2012electrochemical}. \\

Several additives have been added to systems with CB particles to improve their performance\cite{akuzum2020percolation}. Nonionic surfactants have shown to be a good candidate to help stabilize CB slurries by dispersing the particles and allowing the slurry to flow easily by lowering its viscosity\cite{gupta2005adsorption,madec2015surfactant}. The hydrophobic group of the surfactant adsorbs to the hydrophobic surface of CB particles, and the remaining hydrophilic group reaches out to the aqueous medium and provides steric hindrance with other adsorbed surfactants leading to particle stabilization\cite{sis2009effect}. Works have shown that in some applications, the addition of surfactants enabled a homogeneous dispersion of particles that led to the optimization of electrochemical performance\cite{porcher2010optimizing}. However, other studies with the addition of a nonionic surfactant (Triton X-100) have shown that below the critical surfactant concentration (concentration at which enough surfactants have been added to the system and surfactants have saturated the surface of CB particles where any additional surfactants will no longer adsorb to the particle's surface) there is no change in their rheological response, showing that the networks formed by interparticle interactions remain the same\cite{das2024surfactant}. Once enough surfactants have been added above the critical surfactant concentration, there is a sharp change in the mechanical system where the viscosity is suddenly reduced, and the CB particle network is completely broken. This disconnection of the particle network also leads to a catastrophic collapse behavior in the sedimentation of CB particles in the slurry electrodes\cite{lee2023surfactant}. While the mechanical response of this system with the addition of surfactants has been previously explored, there is still a knowledge gap in what the electrochemical response of the system is with the addition of nonionic surfactants, which is critical to understand for battery applications. Further, it should be noted this specific formulation challenge extends more generally to other systems important to energy storage and production\cite{berlinger2021multicomponent}.  \\

Studies have explored the electrical percolation threshold and the mechanism of charge transfer via these conductive particle additives, and how the particle dispersions and aggregate sizes caused by either external shear or other additives change the electrochemical response of the system\cite{hipp2024quantifying, narayanan2017mechanical, akuzum2017effects, karasek1996percolation}. It is understood that conductive particles contribute to the overall conductivity of the system once a network of particles form throughout the system, known as electrical percolation\cite{bauhofer2009review}. The point at which electrical percolation occurs and particles contribute to the electrochemical response of the system is dictated by the electronic pathways that the particles form in the system, which depend on several factors such as the concentration of particles, morphology, and aggregate size\cite{hipp2024quantifying, schueler1997agglomeration, narayanan2017mechanical}. The mechanism by which electrical network form is different from mechanical network formation and percolation, as studies have shown that particles can form electrical networks even when mechanical networks have not formed\cite{richards2017clustering}. However, most of the studies have been conducted in either static conditions or in flowing but closed systems, both of which are quite different from conditions at which a flow battery would operate. Conductive particles flow in and out of the reactor cell under normal conditions. Furthermore, there are limited studies on how a slurry electrode formulated with activated CB particles behaves in an conductive, aqueous medium with the addition of nonionic surfactants that adsorb to the conductive particle surface. \\

In this work, the electrochemical performance of a flowing activated CB slurry is characterized. Moreover, the impact of the addition of a nonionic surfactant (Triton X-100) that adsorbs to the surface of CB particles on the electrochemical performance is reported. Several variables play a role in the electrochemical response of a flowing slurry, especially considering that some features may be at unsteady-state. However, understanding how the system responds to those variables in conditions that a battery operates is needed to improve and potentially optimize battery performance. Herein, the overall conductivity and capacitance response of the slurry in the electrochemical cell are analyzed and reported with systematic changes in flow and scan rate. These results help in building understanding of the physical mechanism dictating these phenomena.

\section{Materials and Methods}

\label{sec:headings}
\subsection{Carbon black slurry preparation}

CB slurries were prepared by dispersing activated CB particles (YP-50F, Kuraray) in 1M H$_2$SO$_4$ (Fisher Scientific) and 0.55M ZnSO$_4$ electrolyte (Sigma Aldrich, Zinc Sulfate Heptahydrate). While many developing RFBs used all-iron metals such as FeSO$_4$\cite{petek2015slurry}, iron naturally oxidizes when exposed to air. To avoid those reactions but still have a divalent salt in the system, ZnSO$_4$ was chosen. In the first set, the CB loading was varied from 1-16 wt\%. For the rest of the experiment, the CB loading was fixed at 10 wt\%, and a nonionic surfactant, Triton X-100 (Sigma Aldrich, average MW 625), was added to the system at $\alpha$ (= c$_{surf}$/c$_{CB}$) ranging from 0 to 0.7. All slurries were stirred with a magnetic stir bar in 250 mL glass bottles overnight to ensure consistent sample preparation. 

\subsection{Electrochemical measurements}

\begin{figure*}[ht]
\includegraphics[width=0.8\linewidth, center]{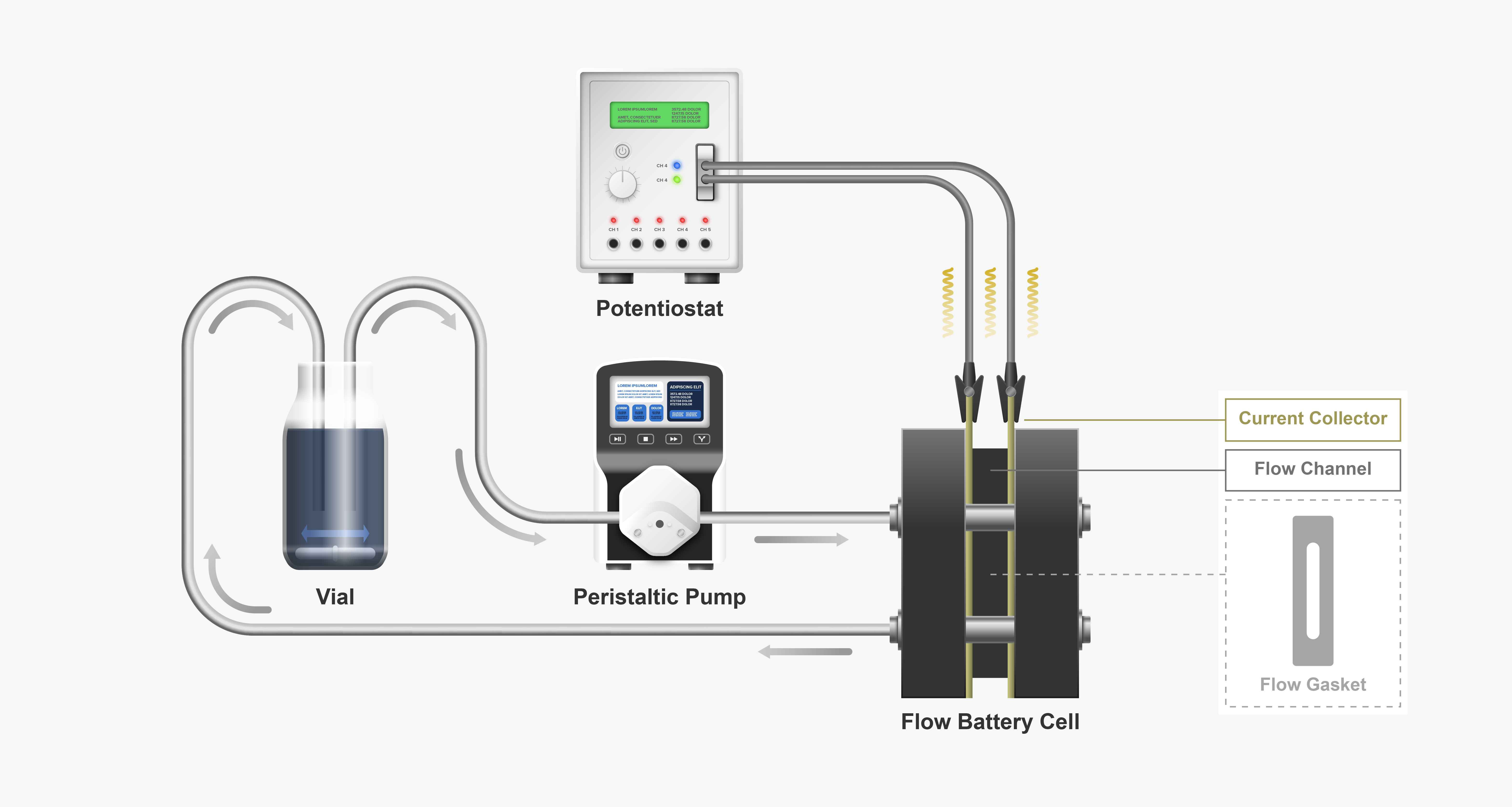}
\caption{\label{fig1}Diagram of the experimental set-up with an unseparated redox flow cell connected to the potentiostat. The peristaltic pump allows slurry in the vial to flow through the cell.}
\end{figure*}

The electrochemical measurements were performed in an unseparated flow battery cell with no separation membrane and a single feed as shown in Fig. \ref{fig1}. A rubber flow gasket  (surface area of 3 cm$^{2}$ and thickness of 0.08 cm) directed the flow of the slurry within the cell. CB slurries were pumped with a peristaltic pump connected through silicon and rubber tubing with ID of 0.25 in at flow rates from 0-170 mL/min. The rubber tubes were only used for the peristaltic pump for mechanical endurance during the measurement. For no flow rate measurements, the slurry was initially pumped at the highest flow rate (170 mL/min) for 1 minute and was left at rest for 15 minutes before starting the measurement.

To measure the capacitance and conductivity of the system, a potentiostat (Reference 600, Gamry Instruments) was used to conduct cyclic voltammetry (CV) measurements over a potential range of $\pm$ 100 mV vs. open circuit potential with scan rates ranging from 1 mV/s to 1 V/s.  The capacitance, \textit{C}, was calculated using the relationship (equation \ref{eq1}):

\begin{equation} \label{eq1}
\frac{\Delta i}{2} = \dot{V}C
\end{equation}

where $\Delta$\textit{i} is the difference between the positive and negative current at a given potential (width of the CV curve), and $\dot{V}$ is the potential scan rate. By plotting $\frac{\Delta i}{2}$ vs. $\dot{V}$, the slope can be calculated for capacitance. The conductivity, $\kappa$ , was calculated using the relationship (equation \ref{eq2}):

\begin{equation}\label{eq2}
\kappa = \dfrac{\partial i}{\partial V}(\dfrac{l}{A})
\end{equation}
where $\dfrac{\partial i}{\partial V}$ is the slope of the CV curve, \textit{l} is the gap between the two current collectors (0.08cm), and \textit{A} is the surface area of the current collector (3cm$^2$). Example calculation can be found in SI section 1 (Fig. S1). 

\par It is essential to note that the experimental apparatus plays an important role in performing these measurements. Although the bottom draw tank is commonly used as a reservoir system\cite{tam2023electrochemical}, conducting long-term measurements caused a significant drop in conductivity of the system due to aggregation of CB particles in the reservoir, thickening the slurry and clogging the narrow channel that the slurry exits through. In this experiment, a 250 mL glass vial was used as the reservoir, and a magnetic stir bar was added for continuous stirring to prevent sedimentation and phase separation that lowers the effective CB loading and decreases the conductivity (Fig. S2). Each experiment was also conducted for < 10 hours to minimize the change in conductivity over time due to the loss of CB particles on the tubing or the vial walls.

\section{Results and Discussion}
\label{sec:others}

\subsection{Effect of carbon black loading}

Two primary components contribute to the overall conductivity of a system consisting of conductive particles dispersed in electrolyte. These include the intrinsic ionic conductivity arising from the solvent electrolyte, and the electronic conductivity resulting from the addition of particles which provide conductive pathways and networks between the current collectors. The addition of particles also increases the capacitance as the particle's surface can form electrical double layers and store charge, and this has been shown to dramatically increase capacitance when a very porous and high surface area particle is used\cite{tam2023electrochemical}. Although electrochemical impedance spectroscopy (EIS) is often used to differentiate components contributing to conductivity and capacitance in a chemical system, conducting such a measurement under flow, with slurry particles having a short, fixed residence time, is challenging. In addition, the heterogeneous nature of the slurry further complicates the measurement. While efforts have shown to use EIS on flowing slurry electrodes, the model is limited and only accounts for the thin surface charge boundary near the current collector surface\cite{hoyt2018electrochemical}. Instead, using CV measurements as the slurry flows through the cell captures the system's overall capacitance and provides an overall picture of the electrochemical response of the slurry. The conductivity reported here from the CV experiment are DC measurements, meaning that only the electronic conductivity coming from the pathways created by the conductive particle additives are measured (see SI section 3 for equivalent circuit verification).

\begin{figure*}[ht]
\includegraphics[width=0.5\linewidth, center]{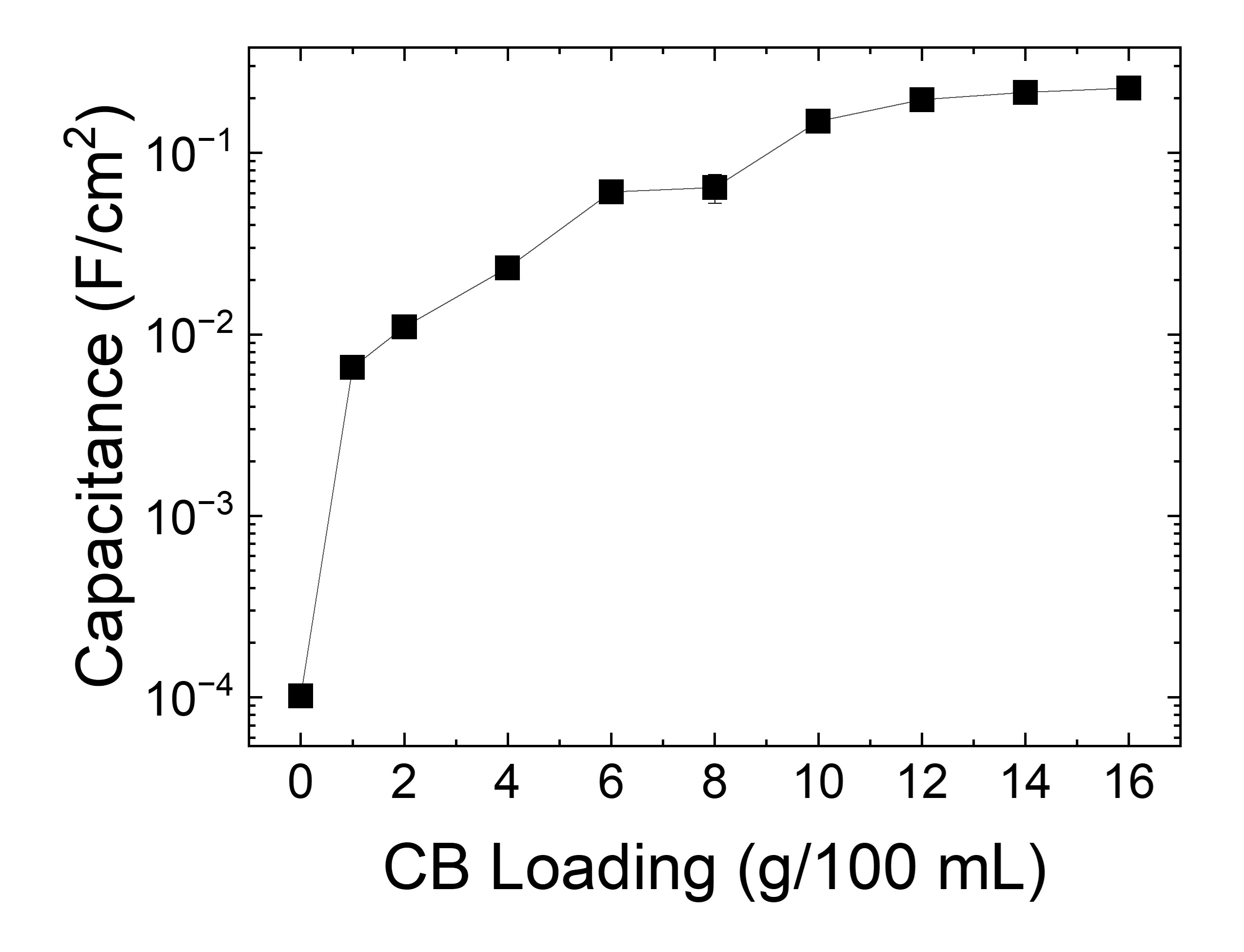} 
\caption{\label{fig2}Capacitance of the CB slurry scaled by the surface area of the current collector with respect to the CB loading (g/100 mL) (flow rate = 0 mL/min and scan rate = 0.1 mV/s)}
\end{figure*}

Capacitance scaled by the surface area of the current collector with respect to the CB slurry loading is reported in Fig. \ref{fig2}. The activated CB used herein (YP-50F) is known to have a very high surface area (1,692 m$^2$/g) and provide more "active sites" and surfaces for electrochemical activity. The magnitude of capacitance primarily depends on the available area for charge to be stored in the electrical double layer. As a result, any addition of CB particles should lead to an increase in capacitance, as additional surfaces of the particles can form electrical double layers and store charge. This is shown with an immediate increase in capacitance even with a small addition of CB where the capacitance increased from 1.0$\times$10$^{-4}$ F/cm$^2$ (with no CB) to 8$\times$10$^{-3}$ F/cm$^2$ at 1 g/100 mL loading.

\begin{figure*}[ht]
\includegraphics[width=0.9\linewidth, center]{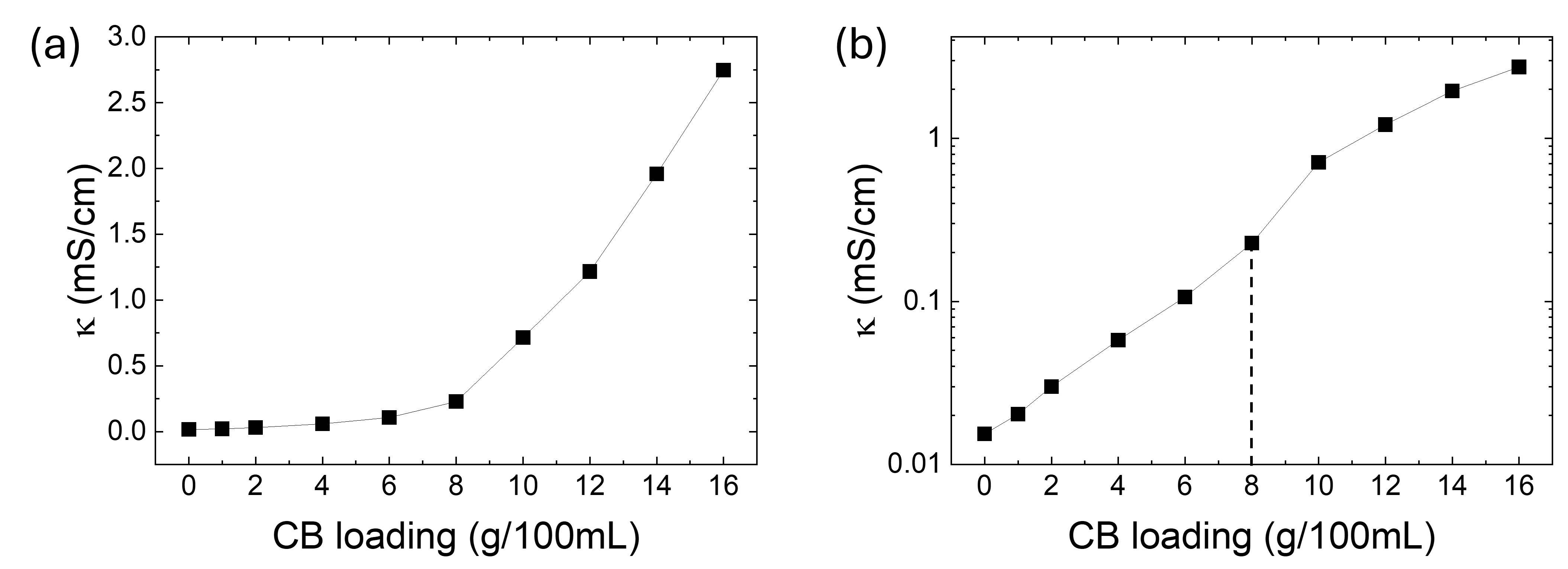} 
\caption{\label{fig3}Conductivity data with respect to the CB loading (g/100 mL) plotted in (a) linear plot and (b) semi-log plot (flow rate = 160 mL/min and scan rate = 100 mV/s). The dotted line indicates the inflection point at around 8 g/100 mL CB loading.}
\end{figure*}

Fig. \ref{fig3} shows the increase in conductivity with respect to CB loading. There are several ways in which the point of electrical percolation can be identified from these data. First, one can plot the conductivity against the CB loading in a semi-log plot (as done in Fig. \ref{fig3}(b)) and locate the inflection point on the graph. The position of this percolation threshold has been found to depend on several factors, including the shape, size, aspect ratio, and dispersion state of the additives and the network it forms based on these factors\cite{tam2023electrochemical,rahaman2017new}. Another strategy for determining the electrical percolation threshold is by identifying the initial point of a strong increase in conductivity compared to the solvent's ionic conductivity (i.e., with no addition of particles), which simply indicates the point at which particles begin to contribute to conduction in the multiphase material. The latter approach suggests a conduction mechanism that does not rely on mechanical contact, rather it is suggestive of a mechanism in which electrons can hop between conductive particle additives. This mechanism would further suggest that electrical percolation can occur at much lower particle concentration as compared to mechanical percolation. These effects would become more pronounced when the continuous phase consists of organic solvents with higher resistance, such as mineral oil or decane\cite{richards2017clustering}.

Using the first convention of identifying an inflection point, one could assess a threshold for electrical percolation of approximately 8 g/100 mL CB loading. However, note that the increase in conductivity is observed even at low CB loadings (1 g/100 mL). The contribution of the CB particle is measured immediately after the conductive particle has been added to the system. The increase in the conductivity shows that CB particles are able to create electronic network and a pathway between the current collectors even at low loadings for charge transfer to happen, meaning that electrical percolation is achieved at a CB loading as low as 1 g/100 mL. With an increase in CB loading, the number of networks and pathways also increase, as shown with the increase in conductivity in Fig. \ref{fig3}. This also explains why the inflection point is not strongly observed in Fig. \ref{fig3}(b). A more pronounced S-shaped curve should be observed when there is a dramatic increase in the conductivity resulting from particle network formation at the electrical percolation threshold. Since the particles already form networks and contribute to the overall conductivity upon addition to the electrolyte, there isn't a single CB loading where the networks form and drastically increase the conductivity. Therefore, the additional contribution of particles to the electrochemical network is less pronounced when compared to systems with conductive particles dispersed in a non-conductive medium\cite{hostettler2023electrical}.

\subsection{Effect of flow and scan rate}

\begin{figure*}[ht]
\includegraphics[width=0.9\linewidth, center]{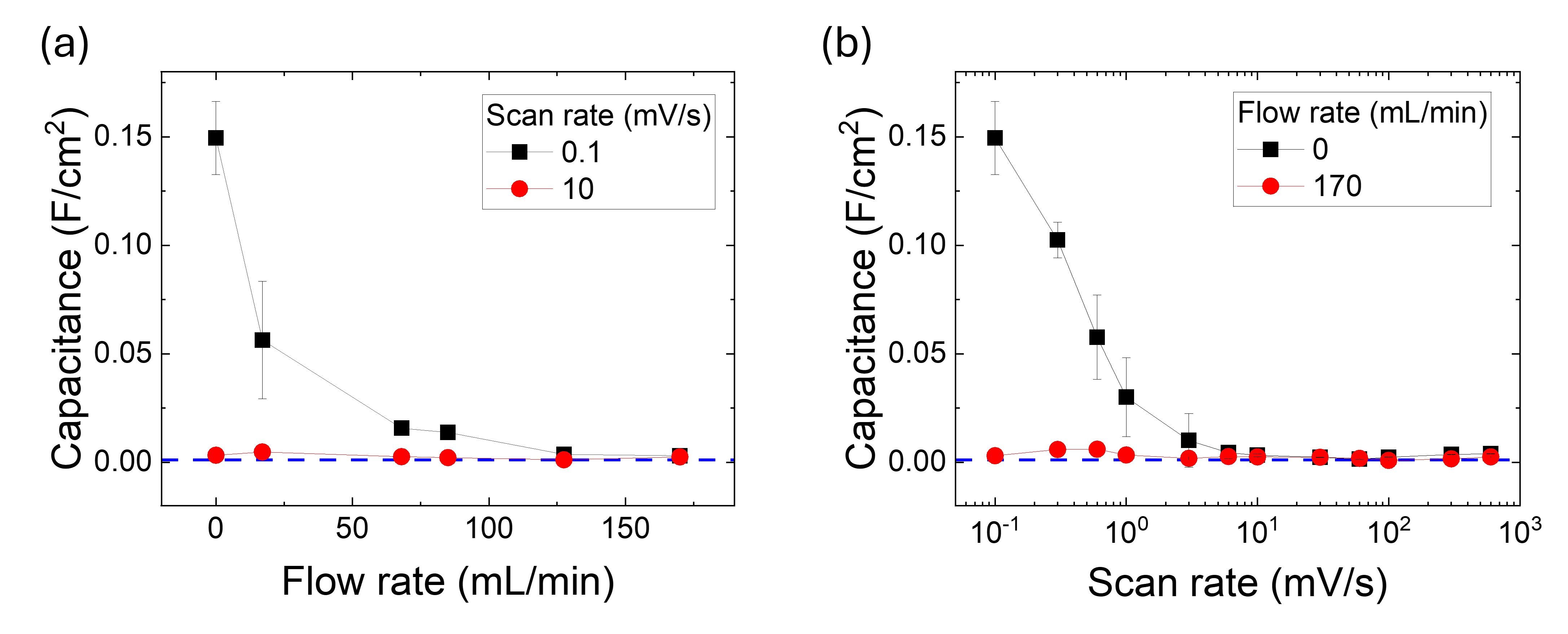} 
\caption{\label{fig4}Capacitance data of CB slurry at 10 g/100 mL loading with respect to (a) flow rate of slurry flowing through the cell and (b) scan rate of the CV measurement. The blue dotted line indicates the measured capacitance of the cell with no CB particles.}
\end{figure*} 

Measurements at systemically varied flow rates and scan rates were conducted at a fixed CB loading of 10 g/100 mL to determine the influence of these parameters on electrical performance. Figures \ref{fig4}(a) and \ref{fig4}(b) show the measured capacitance values with respect to the flow rate of the slurry and the scan rate of the CV measurement, respectively. Fig. \ref{fig4}(a) shows that the capacitance decreases with an increase in the flow rate of the slurry, but only at low scan rates. As the flow rates increase, the capacitance decreases until the measured capacitance becomes similar to that of the value with no CB particles (blue dotted line) at the highest flow rate (170 mL/min). At high flow rates, the particles flow in and out of the cell too fast and the residence time becomes short compared to the time necessary to charge the electrical double layer. Only at low flow rates do the particles remain in the channel long enough for electrical double layers to charge, leading to an increase in the measured capacitance. Therefore, at high flow rates, there is minimal contribution of the particles to the system and agrees observations found in literature \cite{hoyt2018electrochemical, hoyt2015mathematical}.
 
Fig. \ref{fig4}(b) shows that the capacitance also decreases with an increasing scan rate, but only at low flow rates (because particles have to stay in the channel long enough for electrical double layers to charge). Again, the CB particles used here have a high surface area due to the large distribution of micro and mesopores. Since the charging of electrical double layers in the pores is limited by ion transport compared to the outer surfaces of the particles, the electrical double layers cannot charge in the pores at high scan rates ($\ge$ 10 mV/s under no flow conditions). As a result, only the electrical double layers in the outer surfaces of the particles charge, leading to a small or negligible increase in capacitance, as most of the particle's surface area is located in the pores. At low scan rates, the surfaces in the pores can have electrical double layers charge, allowing the particles to contribute to the overall capacitance of the system. Therefore, similar to the effect of flow rate, the contribution of the particles to the system's capacitance can only be measured at low scan rates. Regarding the capacitance contribution of CB particles in the slurry, measurements taken at high flow rates and scan rates (red data points in Fig. \ref{fig4}) limit the contribution of the high surface area particles. This indicates that measurements should be conducted at low flow rates and scan rates to capture the full contribution of CB particles to the capacitance.

\begin{figure*}[ht]
\includegraphics[width=0.9\linewidth, center]{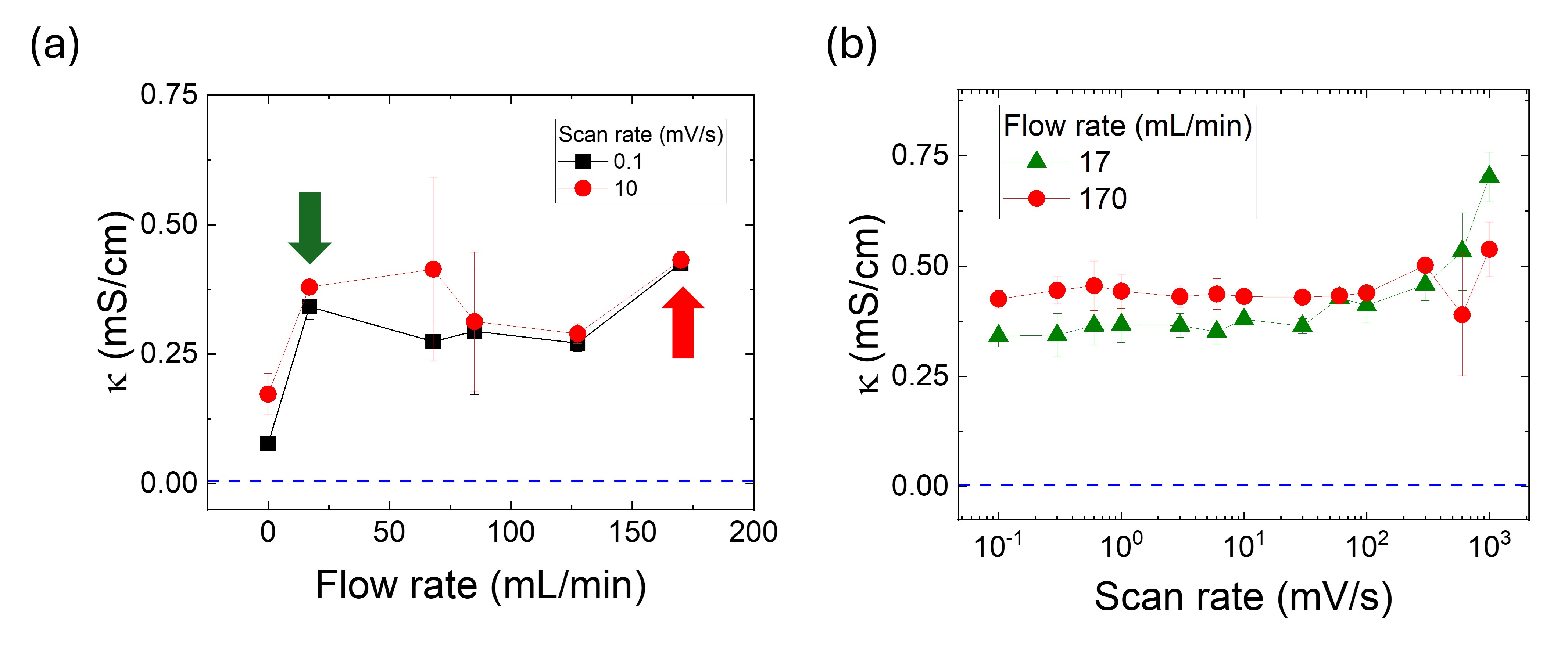} 
\caption{Conductivity of the CB slurry at 10 g/100 mL loading with respect to (a) flow rate of slurry flowing through the cell at 0.1 and 10 mV/s scan rate and (b) scan rate of the CV measurement in flowing conditions. Red and green arrows in (a) correspond to the matching color data points in plot (b). The blue dotted line indicates the measured conductivity with no CB particles.}
\label{fig5}
\end{figure*} 

The change in conductivity at different flow rates is primarily an effect of the shear stress exerted on the sample as it flows through the narrow channel at varying shear rates. High flow and shear rates break up the CB particle aggregates, leaving behind smaller-sized CB particle agglomerates that flow through the channel\cite{richards2017clustering, hipp2021direct}. An increase in the number of these smaller aggregates create more electronic pathways through which charges can transfer and increase the conductivity; however, this is counteracted by the decrease in the number of physical contacts between the particle agglomerates since the aggregates and networks are broken up\cite{narayanan2017mechanical, helal2016simultaneous, amari1990flow}. When the high shear stress break up the particles to smaller sized aggregates, the particles would not be physically connected as would be done in no shear/flow conditions. Still, works have shown that the mode of charge transfer through electronic conductive pathway happens with electron tunneling or hopping, where particles in close proximity can still transfer charge, leading to overall increase in the conductivity\cite{youssry2013non, guy2006critical, awarke2011percolation, richards2017clustering}.

As shown in Fig. \ref{fig5}(a), there is a sudden increase in the conductivity from $\le$ 0.17 mS/cm at 0 flow rate to $\sim$0.35 mS/cm immediately upon flow (17 mL/min). Comparing the static condition to the flowing condition shows that the increase in conductivity from the higher number of pathways (created from smaller CB particle aggregates) dominate over the decrease in conductivity from decrease in the number of physical particle contact. However, as the flow rate increases from 17 - 170 mL/min (calculated equivalent shear rate of 70 - 700 s$^{-1}$ based on the geometry of the channel) the conductivity remains relatively unchanged. The changes in the viscosity of the slurry at the corresponding shear rates observed from flow curves indicate that the particle agglomerate sizes are different from the changes in measured viscosity (Fig. S5). Yet, the slurry maintaining similar conductivity at these flow rates suggest that the two effects are balanced, with the increase in conductivity from more pathways being similar to the decrease in conductivity from the gaps created between the particles. Fig. \ref{fig5}(b) shows the conductivity at different scan rates in flowing conditions (data from the flow rate indicated by the same color arrow in Fig. \ref{fig5}(a)). Under flowing conditions, the conductivity is independent of the scan rate and remains relatively constant. 

\subsection{Impact of added surfactant}

The concentration of nonionic surfactant (Triton X-100) was varied from $\alpha$ = 0 - 0.7 in slurries, with the resulting capacitance and conductivity measured. Previous studies have shown that when surfactants are added to the slurry, the hydrophobic head of the surfactants adsorbs to the surface of CB particles until surfactants have saturated its surface at $\alpha$ = 0.7 (referred as the critical $\alpha$)\cite{lee2023surfactant}. While the slurry forms a gel and their mechanical response little changed even with the addition of surfactants at $\alpha$ < 0.7, a sudden step change occurs at $\alpha$ $\ge$ 0.7 where the CB particle's mechanical network breaks, which leads to catastrophic collapse behavior against gravitational stress\cite{lee2023surfactant, das2024surfactant}.

\begin{figure*}[ht]
\includegraphics[width=0.6\linewidth, center]{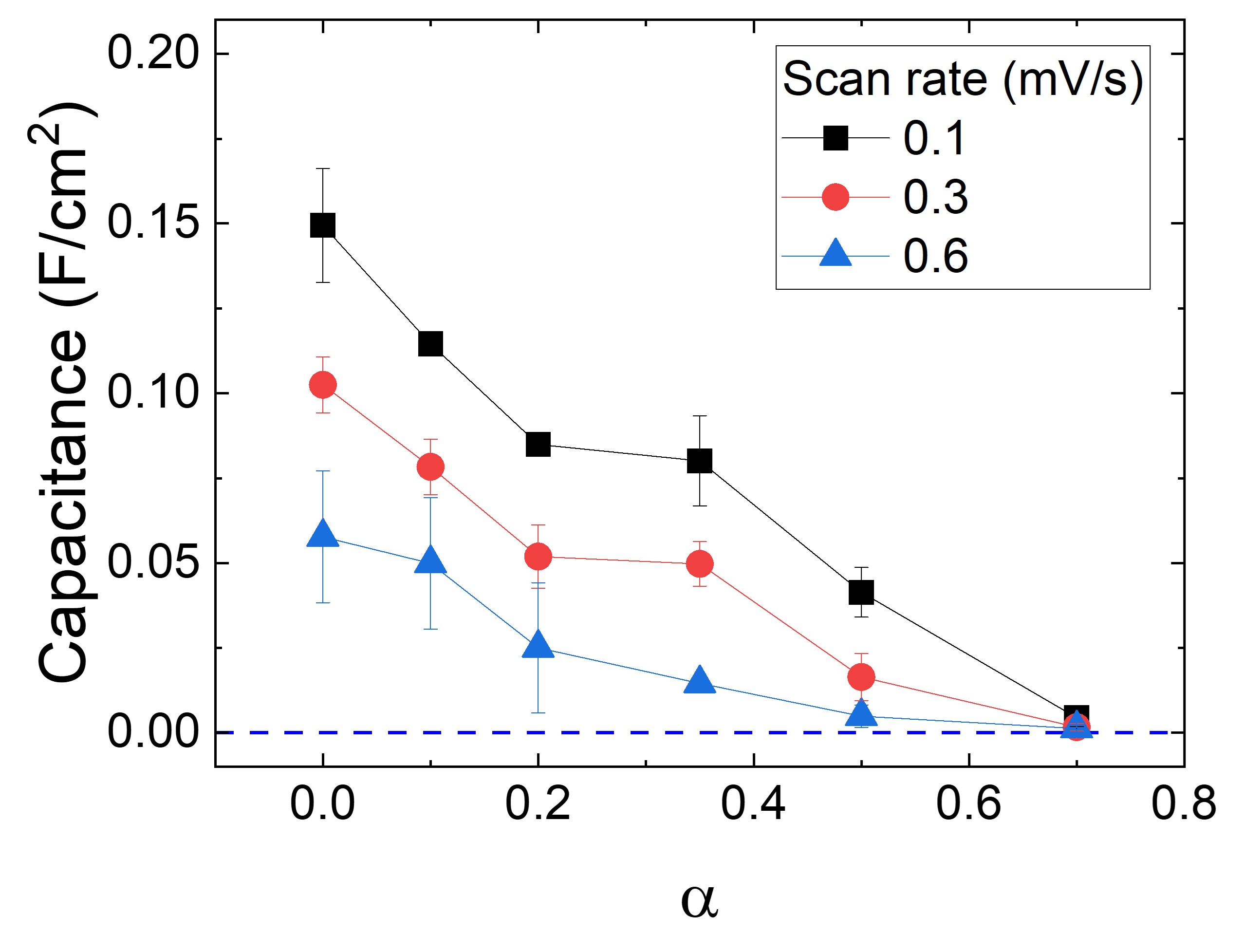} 
\caption{\label{fig6}The gradual decrease in capacitance of the slurry with increasing surfactant concentration, $\alpha$, from 0 to 0.7 at low scan rates (0.1, 0.3, and 0.6 mV/s). The blue dotted line indicates the measured capacitance of the cell with no CB particles.}
\end{figure*} 

Fig. \ref{fig6} shows the changes in capacitance with respect to $\alpha$ at low scan rates under no-flow conditions. Again, low scan rate (< 1 mV/s) data are reported because the electrical double layers in the pores can only be measured at these scan rates. Fig. \ref{fig6} shows that the capacitance gradually decreases with the addition of surfactants until $\alpha$ = 0.7 at which the capacitance drops near to a level similar to the condition without CB particles in the slurry (blue dotted line). Particles lose charge when nonionic surfactants adsorb onto the particle surface, as the surfactants disrupt the charging of the electrical double layer\cite{sis2009effect, somasundaran1992effect}. Since Triton X-100 is a nonionic surfactant that can neither conduct nor store charge, adsorbed particles will not have any alternate sites for the formation of electrical double layers, resulting in the loss of CB particles' ability to store charge and contribute to the system's overall capacitance. When surfactants are added at concentrations where $\alpha$ < 0.7, some free surfaces remain unoccupied by surfactants, allowing electrical double layers to still charge. However, as more surfactants are added to the slurry, the available area gradually decreases, leading to a gradual decrease in the overall capacitance. When surfactants have saturated the surface of CB particles at $\alpha$ = 0.7, there is no space for electrical double layers to charge, effectively removing the particle's contribution to the system's capacitance and its ability to store charge. Although CB particles are still dispersed in the system, surfactants have effectively removed the particles from contributing to the system's overall capacitance.

\begin{figure*}[ht]
\includegraphics[width=0.7  \linewidth, center]{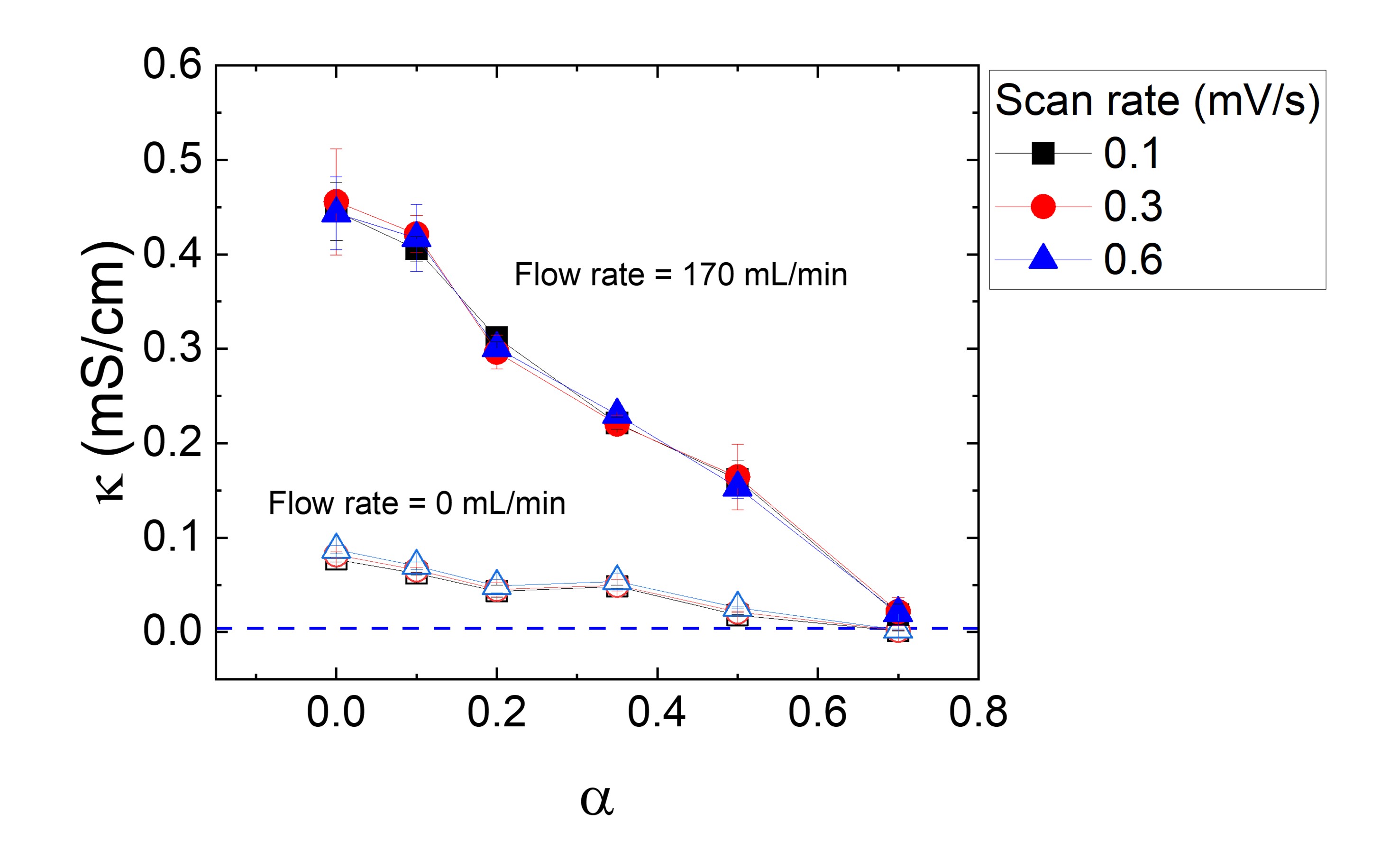} 
\caption{\label{fig7}Conductivity of the slurry at different surfactant concentration, $\alpha$ at low scan rates (0.1, 0.3, and 0.6 mV/s) for slurry flowing at 170 (filled) and 0 (open) mL/min. In all scan rates reported, the conductivity gradually decreases with increasing $\alpha$ until $\alpha$ = 0.7, where the conductivity drops near the conductivity without any CB particles at the same experimental conditions (blue dotted line).}
\end{figure*} 

Similarly, Fig. \ref{fig7} shows that conductivity gradually decreases with increasing surfactant concentration under both static (open) and flowing (filled) conditions, although the slurry flowing at 170 mL/min exhibits higher conductivity than the static case. Considering the static conditions first, rheological studies discussed in earlier work showed that the particles can mechanically percolate and form networks, and that the percolated networks remain mostly undisturbed with addition of surfactants until $\alpha = 0.7$\cite{das2024surfactant}. However, at these static conditions, data shows an immediate decrease in conductivity upon surfactant addition, even at $\alpha = 0.1$, indicating disruption of the electrical pathways. With further addition of surfactants, the conductivity gradually decreases until $\alpha = 0.7$ where the conductivity nears that of the solvent with no CB particles (blue dotted line) and the contribution of particles to the electronic conductive pathway is completely lost. 

It is important to note the profound difference between the electrical and mechanical responses, which suggest that the mechanisms underlying each are distinct, and that disruption of percolation in one may not necessarily correlate with disruption in the other. Again, the surfactant used here (Triton X-100) is a nonionic, non-conductive surfactant, and does not transfer charge. Thus, adsorption of surfactants to the surface of the particle can effectively remove the particle from contributing to the electronic pathway. This would lead to immediate decrease in conductivity as the total number of electronic pathways would decrease. However, this may not disrupt the strength of the mechanically percolated gel as particles with adsorbed polymers may still alow particle flocculation through bridging interactions at low polymer concentration\cite{dickinson1991particle, lafuma1991bridging, otsubo1990elastic, adachi1995dynamic}. In such cases, as the polymer concentration increases and saturates the particle surface, the system can transition to a sterically stabilized phase\cite{dickinson1991particle, adachi1995dynamic}. This provides a plausible explanation for the drastic change observed in the mechanical network at surfactant saturation. The strength of the mechanically percolated network, where some connections are maintained through polymer (surfactant) bridging, may vary with the polymer length. For further insight, additional rheological measurements, such as aging, can be performed in the future.

Next, considering the conductivity data for flowing slurries (filled data points), the conductivity also gradually decreases with addition of surfactants. Here, unlike mechanically percolated slurries, physical networks are broken up under flow, and electronic conductive pathways are formed through smaller CB particle aggregates which current can flow through by tunneling or hopping of electrons\cite{youssry2013non, guy2006critical, awarke2011percolation, richards2017clustering}. The adsorption of non-conductive surfactants onto particle surfaces inhibits electron hopping or tunneling between neighboring particles. Under flowing conditions, electronic conductive pathways are temporary, constantly being created and destroyed simply from flow. However, in regions where surfactants are adsorbed, the formation of such conductive pathways is suppressed. In contrast, surfactant-free particles and surfaces retain the ability to form these pathways. This results in a macroscopic decrease in overall conductivity due to the reduced number of electronically conductive pathways. In both static and flowing cases, the thickness of the surfactant layer adsorbed will impact the formation of electronic conductive pathways, as studies have shown that conductivity is exponentially proportional to the thickness of the insulating polymer layer\cite{guy2006critical, lestriez2010functions}. Thus, performing the CV measurements with the other nonionic surfactants could provide further insight on the surfactants ability to inhibit the electron tunneling and formation of the conductive pathways.

\section{Conclusion}

This work highlights how experimental conditions, such as flow rate and scan rate, are critical in measuring conductive particles' contribution to the overall electrochemical performance of CB slurry electrolyte used in RFB applications. From these data, we conclude that upon addition of conductive particles, they contribute to the electronic conductive pathway, showing immediate increase in the conductivity even at low CB loadings. Regarding the particle's capacitive contribution, data shows that the capacitance increases with addition of CB particles, but the full contribution of particles to the overall capacitance can only be measured at low flow rates and scan rates. The conductivity data shows that the flow rate of the slurry impacts the conductivity as the measured conductivity of a flowing slurry is higher compared to that of a static slurry. Finally, from these data, we conclude that both capacitance and conductivity gradually decrease with the addition of surfactants. When surfactants adsorb to the CB particles, it effectively limits the particle's contribution to both the conductivity and capacitance of the system. 

Electrochemical data with addition of surfactants showed a profound difference in the system's response with addition of surfactants. A step-change in the mechanical system was observed in both sedimentation and rheological experiments at the critical $\alpha$ at which the particle networks no longer charge and the gel-like system breaks down\cite{lee2023surfactant, das2024surfactant}. In contrast, the electrical response is gradual until the critical $\alpha$, where the particles lose their ability to store and conduct charge, resulting in a loss of conductivity and capacitance in the slurry. Overall, this work describes the fundamental behavior and the mechanism in which conductive particles in slurry electrodes behave and their electrical response to additional additives such as surfactants. Further, this work can support formulation decisions and optimizations for slurry electrodes used in RFB applications. 

\section{\label{sec:5}Acknowledgments}
\par The authors acknowledge the financial support from Department of Energy, Office of Electricity, Pacific Northwest National Laboratory, USA (Contract No. 540358). The authors thank Vincent Tam (Case Western Reserve University) for technical discussions.

\printbibliography


\section*{Supplementary material}

# Impact of Surfactant and Flow Rate on the Electrical Properties of Activated Carbon Black Suspensions

## Supporting Information

KangJin Lee, Jesse S. Wainright, and Christopher L. Wirth

May 20, 2025

## 1 Example conductivity and capacitance calculation from CV curve

Data from a CV experiment for a 10 g/100 mL carbon black loading sample with no surfactant measured at 0 flow rate and 10 mV/s scan rate is shown below in Fig. S1. Fig. S1(a) shows the linear fitting to calculate the slope $\frac{\partial i}{\partial V}$ which is then multiplied by $\frac{l}{A}$ where $l$ is the gap between the current collectors (0.08cm) and $A$ is the surface area of the current collector (3cm$^2$). Fig. S1(b) shows how $\Delta i$ was obtained from the CV curve to calculate capacitance using the equation: $\frac{\Delta i}{2} = \dot{V}C$. The capacitance value was obtained by $\frac{\Delta i}{2}$ against $\dot{V}$, and calculating the slope of the curve at a given $\dot{V}$.

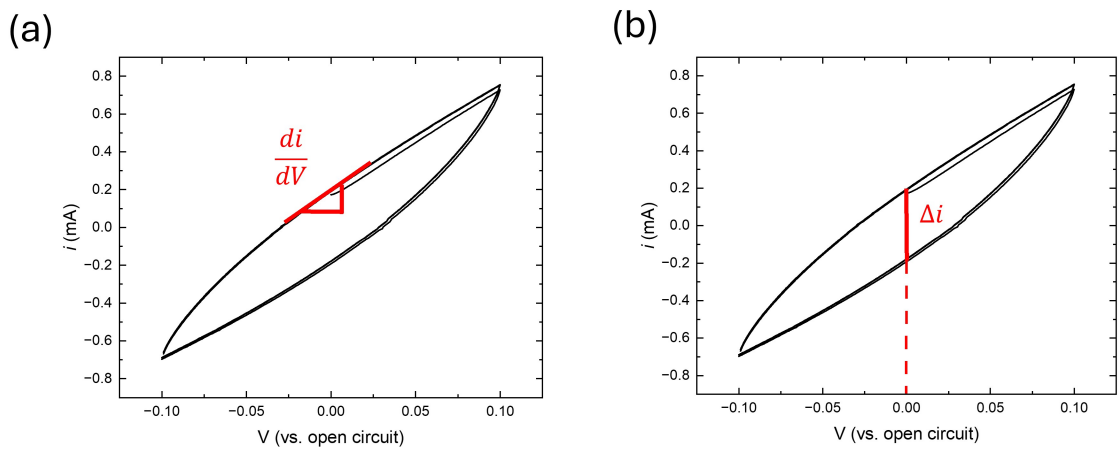

Figure S1: Example data from a CV experiment for a 10 g/100 mL carbon black loading sample with no surfactant measured at 0 flow rate and 10 mV/s scan rate. (a) shows the linear fitting to calculate the slope for conductivity and (b) shows the $\Delta i$ which is then used to calculate the capacitance.

## 2 Experimental set-up

Experimental condition optimization was necessary for electrochemical measurements to ensure that measurements were repeatable and accurate during the course of the experiment. Figure S2 shows the conductivity measured over time for a 10 g/100 mL CB loading slurry flowing at 170 mL/min with and without stirring. First, there is an inevitable decrease in conductivity over time for all the measurements due to loss of CB particles either in tubing walls or in the reservoir container. To mitigate such effects, each electrochemical measurements reported in the main text were limited to 10 hour experiments.

Next, the necessity of incorporating stirring of the slurry in the reservoir is measured when comparing the conductivity with and without stirring. (black and blue curves in Figure S2). When the slurry was continually stirred, there was no drastic change in the conductivity over time as observed when stirring was stopped during the experiment. When stirring was resumed the following day, the conductivity increased again, showing that keeping the slurry continually stirred throughout the experiment is important to minimize the effect of the slurry instability in the reservoir.

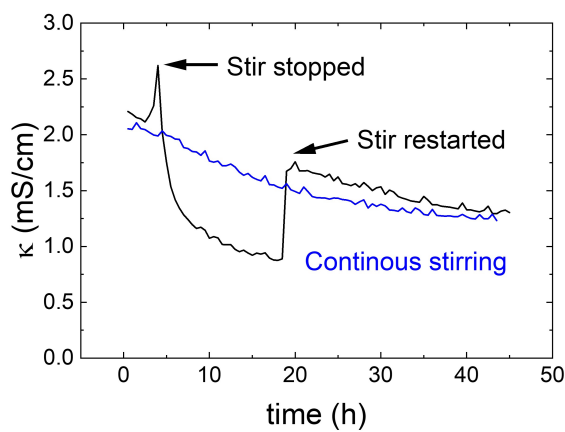

Figure S2: Conductivity of the slurry (10 g/100 mL loading) over time for with stirring (blue) without stirring in the middle of the experiment (black).

## 3 Spice circuit simulation

Spice software was used to create equivalent circuits that simply and represent two conditions of carbon black slurry (above and below percolation) being characterized with cyclic voltammetry (CV) experiment.

Fig. S3 (a) shows the equivalent circuit where both the electronic and ionic conductive pathway is present in the system. This represents the condition at which enough particles are added above the electrical percolation where particles create an electronic pathway between the current collectors in addition to the ionic pathway present from the aqueous medium. C1 mimics the double layer capacitance at the current collectors and R1, R2 and R3 represent the electronic conduction path through the slurry. For the cell's aspect ratio, an electronic conductivity of 0.001 S/cm is equivalent to a resistance of 30 ohm, here shown as three 10 ohm segments. C2 and C3 represent the double layer capacitance of the slurry, consistent with the values reported in the text. R4, R5 and R6 represent the ionic conduction path. The values are based on 1M $H_2SO_4$ with a conductivity of 0.35 S/cm and the cell's aspect ratio, again the total resistance is split into three segments.

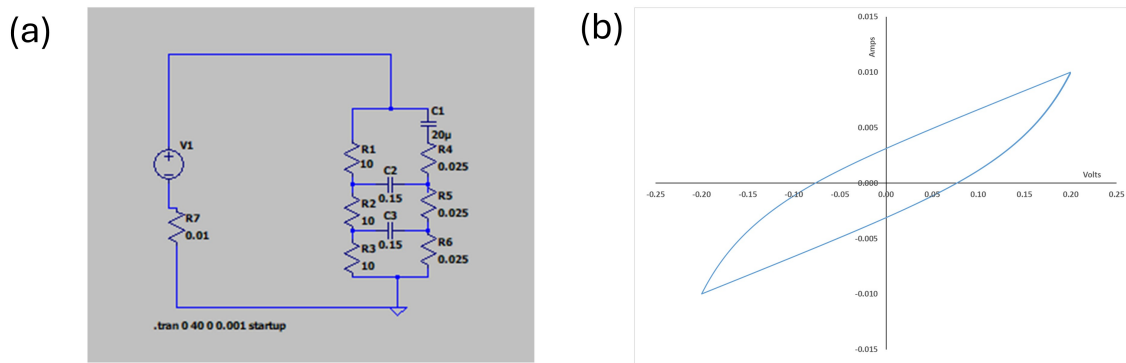

Figure S3: (a) Equivalent circuit modeled for a carbon black slurry above the electrical percolation limit where both the ionic and electronic conductive pathways are present (b) Result of a CV experiment using the equivalent circuit shown in (a) at 40 mV/s between +/- 0.2V.

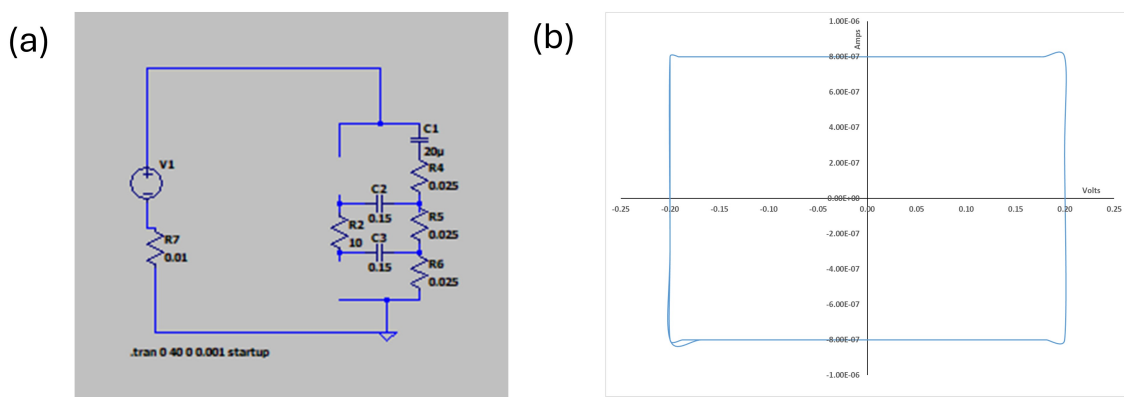

Figure S4: (a) Equivalent circuit modeled for a carbon black slurry below the electrical percolation limit where the electronic connection is removed to mimic the isolated particle in the bulk (ionic pathway still present) (b) Result of a CV experiment using the equivalent circuit shown in (a) at 40 mV/s between +/- 0.2V.

Using the equivalent circuit, CV was simulated at 40 mV/s between +/- 0.2V, producing the result (CV curve) seen in Fig. S3 (b) where the inverse of the slope in the CV is exactly 30 ohms = R1 + R2 + R3.

Then, another equivalent circuit where the electronic connection is removed was simulated as shown in Fig. S4. This mimics the carbon black slurry below the electrical percolation where a particle is isolated between the current collectors in the bulk. The other values are kept the same as equivalent circuit shown in Fig. S3 (a) and CV was simulated at 40 mV/s between +/- 0.2V.Fig. S4 (b) shows the results where the slope becomes flat (no electronic pathway resistant measured) and the capacitance derived is 0.02 mF – the capacitance at the current collectors.

# 4 Rheological measurements of CB slurry

The slurry viscosity was measured at shear rates matching the equivalent shear rate that the slurry flows through the electrochemical cell (70 - 700 $s^{-1}$). Plate-plate rough geometry was used (diameter 25mm) The

changes in the viscosity indicate that the agglomerate sizes are different in each shear rate due to the shear stress exerted on the particles that break the network.

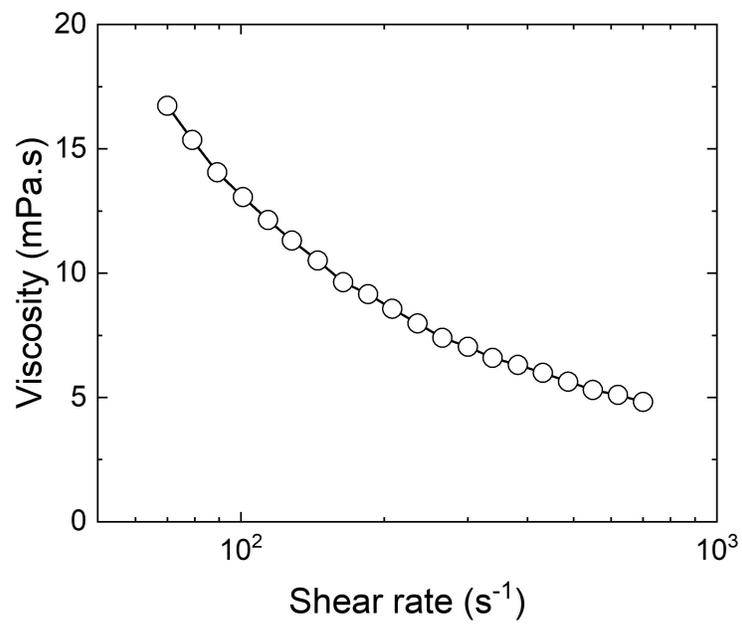

Figure S5: Viscosity of 10 g/100 mL carbon black loading slurry measured at shear rate from 70 - 700 s$^{-1}$.


\end{document}